\documentclass[letterpaper,english,prl,notitlepage,twocolumn]{revtex4-2}
\usepackage{verbatim}
\usepackage{amsmath}
\usepackage{amssymb}
\usepackage{graphicx}
\usepackage{dsfont}
\usepackage{soul}

\makeatletter

\usepackage{color}
\usepackage[usenames,dvipsnames,svgnames,table]{xcolor}
\usepackage{graphicx}
\usepackage{comment}
\usepackage{ulem}
\definecolor{violet}{rgb}{0.58, 0.0, 0.83}

\newcommand{\gae}{\lower 2pt \hbox{$\,
\buildrel{\scriptstyle >}\over {\scriptstyle \sim}\,$}}
\newcommand{\lae}{\lower 2pt \hbox{$\,
\buildrel{\scriptstyle <}\over {\scriptstyle \sim}\,$}}

\makeatother

\begin{document}
\title{Anomalous Floquet-Anderson Insulator with Quasiperiodic Temporal Noise}
\author{Peng Peng Zheng, Christopher I. Timms, Michael H. Kolodrubetz} 
\affiliation{Department of Physics, University of Texas at Dallas, Richardson, TX, USA} 

\begin{abstract}
Time-periodic (Floquet) drive can give rise to novel symmetry breaking and topological phases of matter. Recently, we showed that a quintessential Floquet topological phase known as the anomalous Floquet-Anderson insulator is stable to noise on the timing of its Floquet drive. Here, we perturb the anomalous Floquet-Anderson insulator at a single incommensurate frequency, resulting in a quasiperiodic 2-tone drive. Our numerics indicate that a robust topological phase survives at weak noise with topological pumping that is more stable than the case of white noise. Within the topological phase, we show that particles move subdiffusively, which is directly responsible for stabilizing topological transport. Surprisingly, we discover that when quasiperiodic noise is sufficiently strong to kill topology, the system appears to exhibit diffusive dynamics, suggesting that the correlated structure of the quasiperiodic noise becomes irrelevant.
\end{abstract}
\maketitle

\textit{Introduction} -- Time-dependent Hamiltonians exhibit a wide variety of quantum phenomena that cannot be found in static systems  \cite{1a}. Periodic time-dependence, a.k.a. Floquet drive, for instance opens up the possibility of discrete time crystals \cite{Khemani2016,MBLTOPO3,Else2016a,DFTC3,zhang2017observation,Else2020,KhemaniArxiv2019,Mi2022} and Floquet symmetry-protected topological states \cite{Rudner2013,Roy2016,MBLTOPO2,Else2016,Potter2016,Nathan2015,Roy2017,Titum2016,Po2016,Potter2017,Harper2017,Roy2017a,Po2017,Potter2018,Reiss2018,Kolodrubetz2018,Nathan2019,Timms2021,Long2021,Nathan2021}. A canonical example of Floquet topology is the anomalous Floquet-Anderson insulator (AFAI), in which the insulating bulk has particles with non-trivial micromotion, even though they are stroboscopically localizable. Despite Anderson localization of the entire bulk, and thus a vanishing Chern number, the AFAI nevertheless has chiral edge states which remain robust to finite disorder. 

More recently, we shown that the AFAI remains stable in the presence of white noise, which breaks the time-periodicity of the Floquet problem. Not only is there a well-defined topological invariant in the presence of noise, but also a quantized topological response -- current pumped by the edge states -- remains precisely quantized up to a time scale set by noise-induced diffusion \cite{5a}. While other papers have suggested formal definitions of topological invariants in open quantum systems via extensions of the Berry phase \cite{6a,6b}, our work suggests that Floquet systems provide unique robustness to noise due to the fact that topology is defined via time evolution of the entire manifold of states, rather than a gapped ground state. 
 
Given that the ultimate loss of topological response is due to noise-induced diffusion, we might ask in which ways the system can be made controllably less noisy in hopes of sustaining the topological response for longer, perhaps indefinitely. If we think of noise as originating from a bath of oscillators with continuous spectrum, we can slowly approach the noisy limit by considering single or few-tone noise sources in addition to the strong Floquet drive. The problem that emerges is one of multi-frequency driving, also known as quasiperiodic Floquet drive. Quasiperiodic drive is interesting in its own right, enabling novel topological phases of matter, symmetry breaking, and other extensions of Floquet theory \cite{7a8a9a,7b8b9b,7c8c9c,Dumitrescu2018,Friedman2022}. Interestingly, quasiperiodic drive has also been argued to be consistent with many-body localization \cite{10a}, which would enable these phases of matter to remain robust to infinite time even in the presence of interactions.
 
In this paper, we consider the simplest case of replacing the white noise from the bath by a single incommensurate drive. We explore topological response in this bichromatically driven system and find finite-time quantization similar to the noisy case. Unlike the noisy case, quasiperiodic drive creates an unusual potential landscape in the frequency lattice, which changes diffusive transport into subdiffusive over a wide range of parameters. Subdiffusion extends the time scale of topologically quantized pumping, though it still eventually decays on time scales set by a power law of system size. In other models, one may be able to find regimes with localization in both spatial and frequency directions, for which the topological response would remain quantized to exponential time in system size.
 
\begin{figure}
\centering
\includegraphics[trim={3cm 3cm 11cm 3cm},width=1.1\columnwidth]{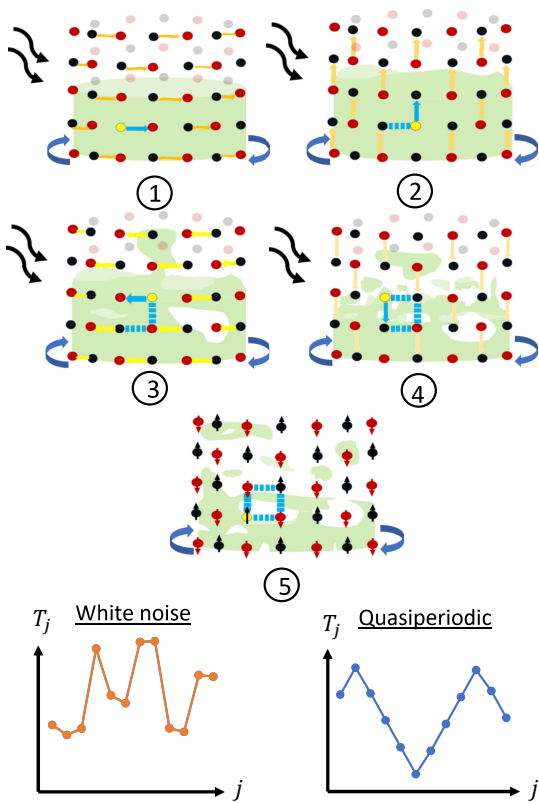}
\caption{Schematic representation of the five-step drive producing the quasiperiodically driven anomalous Floquet-Anderson insulator. In the absence of noise or disorder, bulk fermions hop around one plaquette during each cycle, returning to their original location (light blue arrows). Initially the bottom half of the cylinder is filled, but as time progresses, particles delocalize subdiffusively. Bottom panels illustrate white noise and triangular-wave quasiperiodic noise.}
\label{f:Cylinder}
\end{figure}

 \textit{Model} -- We begin with the conventional model of the AFAI, which consists of spinless fermions hopping on a two-dimensional square lattice. In the absence of noise or quasiperiodic drive, the Hamiltonian is given by a five step drive with period $T$, such that each step has period $T/5$. The first four steps involve hopping between select pairs of sites,
 \begin{equation}
     H_\ell=-J\sum_{\langle ij \rangle_\ell} c_i^{\dagger}c_j
 \end{equation}
 for $\ell \in \{1,2,3,4\}$ where the connected sites $\langle ij \rangle_\ell$ are chosen such that the particles hop in a closed loop around the plaquette each Floquet cycle, as shown in Figure \ref{f:Cylinder}. The hopping strength, $J$, is fine-tuned to $JT=5\pi/2$, such that the particles hop exactly one site per step in the absensce of noise or disorder. The final step is a staggered chemical potential, 
 \begin{equation}
     H_5 = \Delta\sum_j \eta_j c_j^{\dagger}c_j,
 \end{equation}
 where $\eta_j=+1(-1)$ on the $A (B)$ sublattice. We consider cylindrical geometry, with periodic boundary conditions in the $x$ direction and open boundary conditions in the $y$ direction. The system size is $L\times L$ lattice sites. Finally, we add static chemical disorder, 
 \begin{equation}
     H_{dis}=\sum \mu_j c_j^{\dagger} c_j
 \end{equation}
 to give Anderson localization. The strength of chemical potential disorder is set by $W_x$, such that $\mu_j$ is uniformly sampled between $[-W_x,W_x]$.

We add quasiperiodic noise to our model by varying the timings of the five steps. Instead of fixed period $T/5$ per step, we use 
\begin{equation}
    T_j=\frac{T}{5} \left( 1 + \delta_j \right)
\end{equation}
where $\delta_j \in [-W_t,W_t]$ is the temporal disorder of strength $W_t$. In this paper, we compare two types of noise: true temporal disorder, i.e., white noise, in which the $\delta_j$ are chosen independently and randomly as in  \cite{ian}, and quasiperiodic ``disorder,'' in which the $\delta_j$ are a periodic function of time that is incommensurate with the original Floquet period $T$. In particular, we choose
\begin{equation}
\delta_{t}=W_{t}f\left(\frac{2\pi \alpha j}{5} +\Phi\right)
\end{equation}
where $f(\theta)=f(\theta+2\pi)$ is a $2\pi$-periodic function, $\alpha$ is an irrational number, which we choose as the golden ratio $\alpha=(1+\sqrt{5})/2$, and $\Phi$ is the initial phase of the drive. While a natural choice for $f$ is a sinusoidal function, we instead choose the triangular wave: $f(-\pi/2 < \theta < \pi/2) = \theta/(\pi/2)$, $f(\pi/2 < \theta < 3 \pi /2) = f(\pi/2 - \theta)$, such that the probability of a given $\delta_j$ will be evenly distributed from $-W_t$ to $W_t$, enabling direct comparison with white noise. For notational simplicity, since we wish to keep the Floquet period $T$ constant, we instead think of this $1+\delta_j$ factor as rescaling coefficients of the Hamiltonian and keep each step at $T/5$. Our results will not depend on this choice.

The topological response of the AFAI is most robustly found by considering a cylinder geometry in which the bottom half of the cylinder is initially filled and the top half is initially empty. Chiral edge states appear at the top and bottom edges, and our initial state ensures  that only one edge state is filled. This edge state carries current around the cylinder at a quantized rate, while the bulk is localized and thus insulating. Hence, we have quantization of the total charge pumped per Floquet cycle ($Q$), which is defined for a given Floquet cycle by
\begin{equation}
    Q=\int_{t_0}^{t_0+\Tilde{T}} dt \langle\psi(t)|I|\psi(t)\rangle
\end{equation}
where $I$ is the current operator, $\Tilde{T}=\sum_\ell T_\ell$ is the Floquet period modified by the temporal noise, and $t_0$ is the start of the Floquet cycle. Our parameters are chosen such that $Q=1$ in the topological phase. We simulate the system identically to the case with white noise; details may be found in \cite{ian}.

\textit{Results} -- A priori, the effect of quasiperiodic temporal disorder on pumped charge ($Q$) is not obvious. On one hand, quasiperiodically driven systems are amenable to localization, allowing the possibility of exponentially long-lived topological pumping. On the other hand, systems subjected to white noise display quantized topological response up to the Thouless time for sufficiently weak disoder, which was argued to result from direct averaging over multiple Floquet cycles \cite{ian}. By contrast, quasiperiodically driven systems do not self-average in this way, and in particular have the potential to give constructive interference of the pumping over multiple cycles that could destroy its quantization.

\begin{figure}
\centering
\includegraphics[width=1\columnwidth]{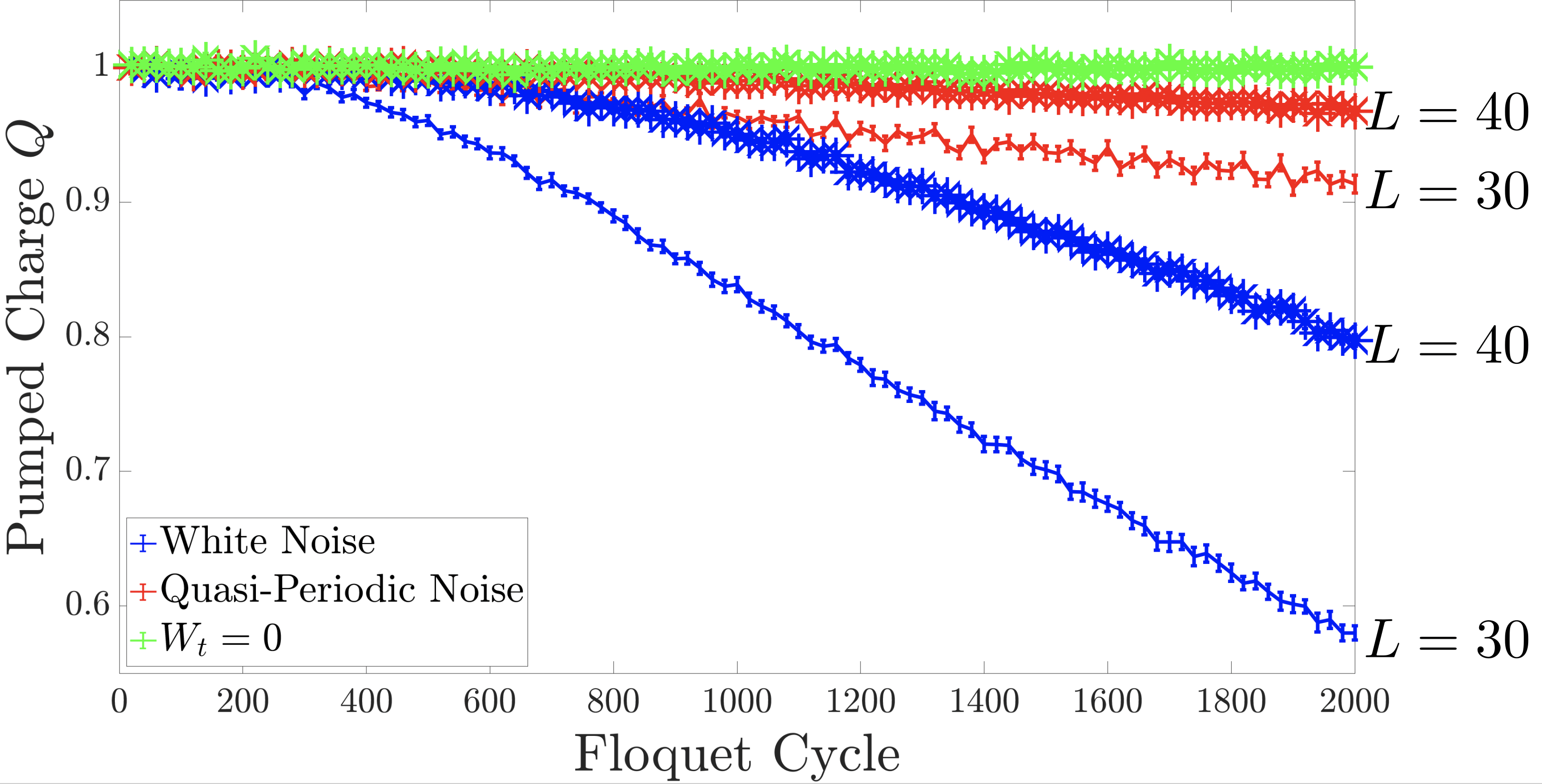}
\caption{Finite size dependence of charge pumping ($Q$) as a function of time with $W_x=1$.
White noise is seen to cause more rapid decay than quasiperiodic noise. The noisy curves have $W_t=0.1$, well within the topological phase of the model with white noise \cite{ian}.}
\label{f:randomvsquasi}
\end{figure}

Therefore, we start with direct simulations of the charge pumped, $Q(t)$, as a function of system size, disorder, and noise strength. A characteristic trace is shown in Figure \ref{f:randomvsquasi} for $W_x=1$ and $W_t=0.1$, which lies in the topological phase of the noisy model \cite{ian}. Both the white noise and quasiperiodic case show the same general trend -- a short-time transient, followed by a long-lived quantized plateau, and finally decay towards $Q=0$. However, we also immediately note two qualitative differences: first, charge pumping in the quasiperiodic case decays much more gradually than the random noise case. Second, in both cases there exists a strong finite size effect, with the decay time $\tau$ sharply increasing with system size. In accordance with the noisy results, we postulate power law scaling of the decay time, $\tau \sim L^\alpha$, and attempt to find the value of $\alpha$ below. 

While quantized charge pumping is the observable consequence of topology, it has strong finite size effects that make it less useful for detecting topological phase transitions \cite{4a16a17a,ian}. Instead, multiple papers have found localization to be a useful metric for detecting the phase transition \cite{4a16a17a,5a,ian}, as the system is localized on both the topological and trivial side of the phase transition. Right at the critical point, however, the system must delocalize in order to rearrange its band and change the topological invariant. Therefore, one can use conventional metrics of (single-particle) delocalization to locate the phase transition in these models.

One conventional metric for localization is the level spacing ratio \cite{19a}. However, that requires identification of eigenstates, which are not easily accessible in our bichromatically driven system. Instead, we consider the participation ratio (PR), which is defined for an arbitrary single-particle state $|\psi\rangle$ and position basis $|\mathbf{r}\rangle$ as
\begin{equation}
    \mathrm{PR} = \left(\sum_\mathbf{r} \left|\langle \mathbf{r} | \psi \rangle\right|^4\right)^{-1}.
\end{equation}
 For a wave function that is fully delocalized in the position basis, such that each basis element occurs with probability $p = 1/L^2$, the PR is equal to the total number of sites, $L^2$. By contrast, if the particle is localized on a single $\mathbf{r}$ site, then $\mathrm{PR}=1$. In general, for a $d$-dimensional system, we can think of $\mathrm{PR}^{1/d}$ as a proxy for the localization length.

\begin{figure*}
\centering
\includegraphics[width=1\textwidth]{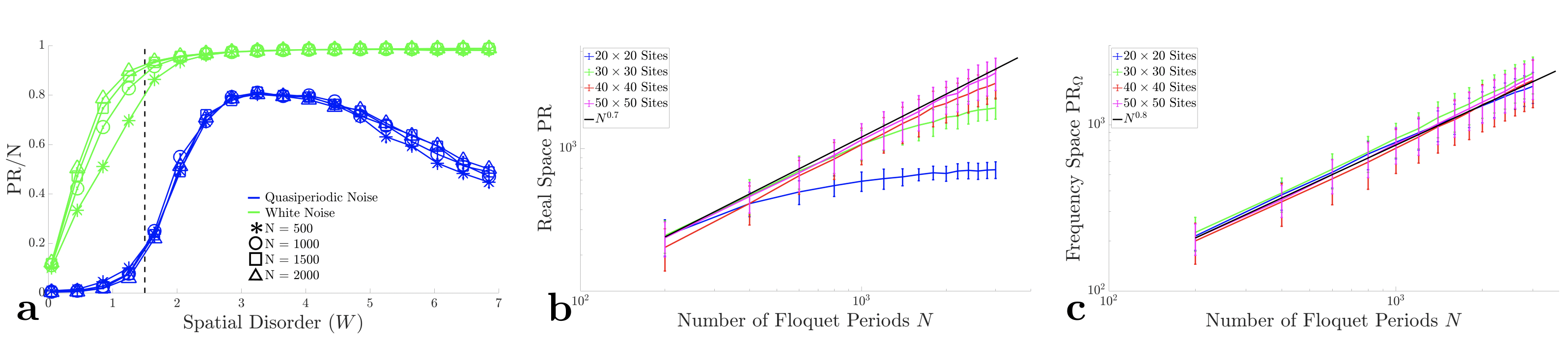}
\caption{Participation ratio (PR) in (a,b) real space and (c) frequency space with temporal disorder $W_t=0.1$. (a) Comparison of white noise and quasiperiodic drive as a function of number of Floquet periods for a system size of $L=30$. In the case of diffusion in two dimensions, we expect $\mathrm{PR} / N \sim \xi^2 / N$ to be independent of $N$, while it will decrease with $N$ in the case of subdiffusion. Note that quasiperiodic drive consistently has smaller PR, indicating stronger localization. (b) Time-dependence of spatial PR and (c) frequency-space PR, showing consistency with subdiffusion in real space and ballistic spreading in frequency space. In (b) and (c), the spatial disorder strength is $W_x=1.5$ as indicated by the dashed black line in (a).}
\label{f:IPRWConst}
\end{figure*}

Here we consider the participation ratio for the wave function $|\psi(N T)\rangle$ obtained by time-evolving an initial state $|\psi(0)\rangle = |\mathbf{r}=(0,0)\rangle$ located in the middle of the system for $N \gg 1$ Floquet cycles.  Participation ratios for various system sizes and noise strengths are shown in Figure \ref{f:IPRWConst}a and b. The first thing we notice is that PR is much smaller for quasiperioidic noise than for white noise, suggesting that, for a given time $NT$, the quasiperiodic system is much more well-localized than the white noise case, which is known to delocalize diffusively. Secondly, we notice strong finite size effects and, more strikingly, large dependence on the evolution time $NT$. Empirically, we see that at large $L$ -- where finite size effects can be neglected -- the PR appears to approach a power law scaling, with $\mathrm{PR} \sim N^{0.70}$ for $W_x=1.5$ and $W_t=0.1$. We will comment further on the exact nature of this power law shortly.

\begin{figure*}
\centering

\includegraphics[width=0.95\textwidth]{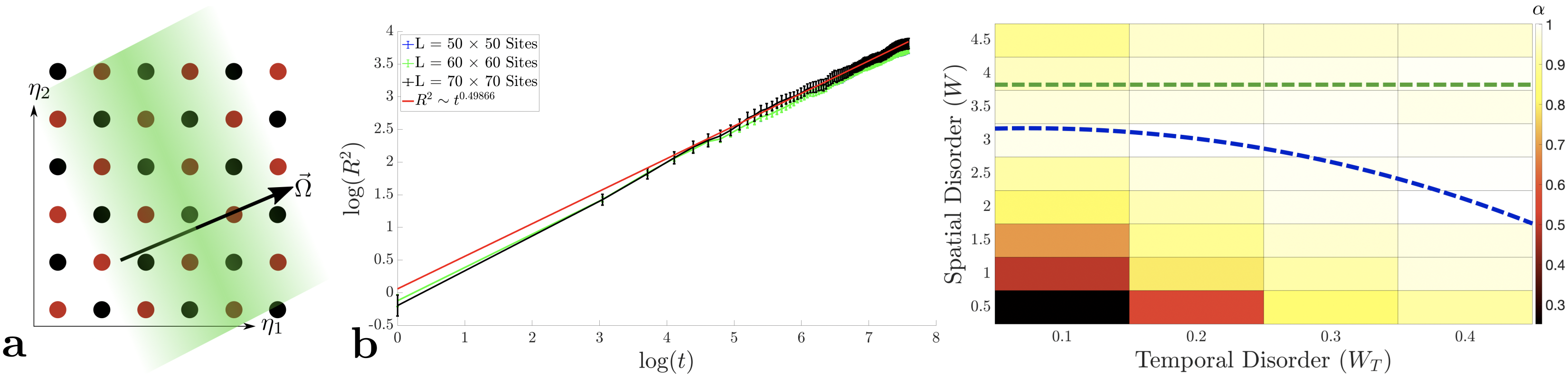}
\caption{(a) Illustration of the Floquet extended zone picture for 2-mode driving. Frequencies $\omega_1$ and $\omega_2$ map to photons, whose wave function is Wannier-Stark localized in the direction parallel to the drive frequency ($\vec{\Omega}$). Localization perpendicular to $\vec{\Omega}$ (green shaded direction) depends on many factors, including irrationality of $\omega_2/\omega_1$. (b) Direct calculation of subdiffusion for $W_x=1$ and $W_t=0.1$. The mean-square radius increases as $\sim t^{0.5}$, well below the diffusive exponent, $t^1$. (c) Dependence of subdiffusive exponent, $R^2 \sim t^\alpha$, on spatial and temporal disorder. Dashed lines show the phase boundaries for white noise, taken from \cite{ian}.}
\label{f:ian}
\end{figure*}

Unlike conventional localization problems, this quasiperiodically driven system also has another axis in which localization must be probed. If we consider mapping the 2-frequency Floquet problem to a photon lattice -- the so-called Floquet extended zone picture -- then the photons themselves live on a two-dimensional square lattice with energy tilt (``electric field'') proportional to the drive frequencies. This extended zone lattice is illustrated in Figure \ref{f:ian}a. While Wannier-Stark localization along the electric field direction is guaranteed, localization in the orthogonal direction (green shaded area) depends on the precise model. In the language of Anderson localization, this orthogonal direction looks similar to quasiperiodic chemical potential disorder via a cut-and-project method \cite{Singh2015}. Taken together, the two spatial dimensions and two frequency dimensions give a single-particle localization problem in a quasi-three-dimensional space, with a combination of uncorrelated (real-space) and quasiperiodic (frequency-space) disorder. We are not aware of a solution to this localization problem. Furthermore, frequency-space localization has been shown to be fundamental in defining topological invariants for other quasiperiodically driven systems \cite{7b8b9b,7c8c9c}. Therefore, frequency space localization will be required for exponentially long-lived topology in our model as well.

We study frequency space localization via a generalized participation ratio, introduced in \cite{7b8b9b}. We start with a discrete Fourier transform of the time-evolved wave function:
\begin{equation}
    |\psi_N(\omega)\rangle = N^{-1/2} \sum_{j=1}^Ne^{i j \omega  T }|\psi(jT)\rangle,
\end{equation}
where $\omega = 2\pi k / (N T)$ with $k=0,1,\ldots,N-1$. If we think of this frequency space lattice as our basis, then the density for ``site'' $\omega$ is given by
\begin{equation}
    \rho_N(\omega) \propto \langle \psi_N(\omega)|\psi_N(\omega)\rangle.
\end{equation}
We choose the normalization $\sum_\omega \rho_N(\omega) = 1$ to match the conventional PR. The frequency space PR is then, by analogy,
\begin{equation}
    \mathrm{PR}_{\Omega} = \left(\sum_\omega \rho_N(\omega)^2\right)^{-1}.
\end{equation}
As with real-space participation ratio, $\mathrm{PR}_\Omega=N$ for the fully delocalized state and $\mathrm{PR}_\Omega=1$ for the fully localized state.

The frequency-space PR is shown in Figure \ref{f:IPRWConst}c. Similar to real-space PR, it shows evidence of delocalization, with strong finite-size and finite-time dependence. A power law fit gives $\mathrm{PR}_\Omega \sim N^{0.8}$ for the parameters shown, which we analyze further below.

\textit{Subdiffusion} -- If we consider $\mathrm{PR}^{1/2}$ and $\mathrm{PR}_\Omega$ as proxies for the real- and frequency-space localization lengths, then the power law fits in Figure \ref{f:IPRWConst} are suggestive of subdiffusive motion. For $W_x=1.5$ and $W_t=0.1$, Figure \ref{f:IPRWConst}b shows spatial $\mathrm{PR} \sim r^2 \sim t^{0.70}$, which is below $r^2 \sim t$ that characterizes diffusion. Frequency-space PR is more complicated, showing PR$_\Omega \sim t^{0.8}$. Since frequency space is effectively one-dimensional due to Wannier-Stark localization along $\vec{\Omega}$, this is superdiffusive but subballistic. Most notably, frequency space and real space do not have the same exponents, indicating that particle delocalization is anisotropic in this extended zone picture.

We can study spatial subdiffusion directly by looking at the variance of the position in the time-evolved state, 
\begin{equation}
    R^2 = \langle \psi(NT) | r^2 | \psi(NT)\rangle.
\end{equation}
Fitting this for a given value of spatial and temporal disorder, as illustrated in Figure \ref{f:ian}b, we find clear subdiffusive power law growth. A phase diagram of the subdiffusive exponent as a function of $W_{x}$ and $W_{t}$ is shown in Figure \ref{f:ian}c. Note that the power laws obtained from $R^2$ and those from PR match within error bars (cf. Figures \ref{f:IPRWConst}b and \ref{f:ian}c). Over the majority of the phase diagram, quasiperiodic drive shows clear differences from white noise, where the dynamics are always diffusive \cite{ian}. For sufficiently strong $W_x$ and $W_t$, the dynamics is not readily distinguished from diffusion, which is consistent with the well-studied possibility of diffusion in three-dimensional disordered systems  \cite{24a,24b}. We have also added curves corresponding to the numerically identified topological phase transitions for the case with white noise, taken from  \cite{ian}. Intriguingly, this transitions align somewhat closely with cases where the exponents appear to approach those of diffusion. One possible origin of this effect is the physics of the crossover phase, which was argued in \cite{ian} to occur for $W_x \lae 3.6$ and $W_t > W_{t,c}(W_x)$ (the phase to the right of the blue line in Figure \ref{f:ian}). In the crossover phase, some realizations of temporal noise correspond to a trivial Floquet problem, while others are topologically non-trivial. As the Anderson localized micromotion within these two phases is topologically incompatible, one may imagine that the cycles add ``incoherently'' despite the structured nature of the quasiperiodic drive. This would effectively restore true decohering noise, and could result in the observed diffusive spreading.

\textit{Discussion} -- We have seen topological pumping in a quasiperiodically driven variant of the anomalous Floquet-Anderson insulator (AFAI). Initially localized states spread subdiffusively, with exponents determined by the strength of spatial disorder ($W_x$) and quasiperiodic temporal disorder ($W_t$). In all situations, the quasiperiodically driven system is seen to be at least as stable as the equivalent noisy system. Given the existence of a stable topological phase with long-lived quantized pumping in the noisy system \cite{ian}, our numerics suggest that such a stable topological phase exists in the quasiperiodically driven system as well.

While it is encouraging that a topological phase persists at finite $W_t$, the nature of the phase and phase transitions remains unclear. It is not exponentially long-lived in system size, as is the case for conventional topological phases of matter that are Anderson localized; despite our best attempts, no fully localized points in the phase diagram were found. In general, the study of combined Wannier-Stark and Anderson localization in systems like ours remains an interesting unsolved problem. Yet we can hope that variants of the model may be found in which localization is present in both spatial and frequency directions, for which a topological invariant could presumably be defined by analogy with the noiseless case. Note that such a topological invariant will involve integrating over the $\omega_2$ direction on the frequency lattice, as the topological pumping only relates to the original Floquet cycle $T=2\pi/\omega_1$. This is similar to a three-dimensional Chern insulator \cite{28a,28b}, suggesting possible extensions to weak and/or higher-order topology \cite{29a,29b,29b30b}. Finally, we note that, were the problem localized in both frequency and position space, it would admit a four-dimensional topological description. There are fewer obvious candidates for (non-symmetry-protected) topology in 4D, as the winding numbers usually used for Floquet topology are only defined in odd dimension  \cite{30a,29b30b}. One possibility is to look for a non-trivial second Chern number \cite{31a,31b}. Whether second Chern insulators can be extended to a bichromatically driven two-dimensional system is an interesting open question.

\textit{Acknowledgments} -- We thank Sarang Gopalakrishnan and Lukas Sieberer for valuable discussions. This work was performed with support from the National Science Foundation through Grant No. DMR-1945529 and the Welch Foundation through Grant No. AT-2036-20200401 (M. K. and R. G.). S. G. acknowledges support from the Israel Science Foundation, Grant No. 1686/18. F. N. and M. R. gratefully acknowledge the support of the European Research Council (ERC) under the European Union Horizon 2020 Research and Innovation Programme (Grant Agreement No. 678862), and the Villum Foundation. We used the computational resources of the Lonestar 5 cluster operated by the Texas Advanced Computing Center at the University of Texas at Austin and the Ganymede and Topo clusters operated by the University of Texas at Dallas’ Cyberinfrastructure \& Research Services Department.

\bibliography{AFAI}

\end{document}